\documentstyle[twocolumn,aps,prl,floats,psfig]{revtex}
\begin{document}
\draft

\voffset=0.8in
\title{PWIA Extraction of the Neutron Magnetic Form Factor  
from Quasi-Elastic $\bf ^3\vec{\bf He}(\vec{e},e')$
at $\bf Q^2 = 0.3$ to 0.6 (GeV/c)$^2$}

\author{W.~Xu,$^{12}$ B.~Anderson,$^{10}$ 
L.~Auberbach,$^{19}$ 
T.~Averett,$^{3}$ W.~Bertozzi,$^{12}$ T.~Black,$^{12}$ J.~Calarco,$^{22}$ 
L.~Cardman,$^{20}$ G.~D.~Cates,$^{15}$ Z.~W.~Chai,$^{12}$
J.~P.~Chen,$^{20}$ S.~Choi,$^{19}$ E.~Chudakov,$^{20}$ 
S.~Churchwell,$^{4}$ G.~S.~Corrado,$^{15}$ C.~Crawford,$^{12}$ 
D.~Dale,$^{21}$ 
A.~Deur,$^{11,20}$ P.~Djawotho,$^{3}$ T.W. Donnelly, $^{12}$
D.~Dutta,$^{12}$ 
J.~M.~Finn,$^{3}$ 
H.~Gao,$^{12}$ R.~Gilman,$^{17,20}$ A.~V.~Glamazdin,$^{9}$ 
C.~Glashausser,$^{17}$ 
W.~Gl\"{o}ckle,$^{16}$ J.~Golak,$^{8}$
J.~Gomez,$^{20}$ V.~G.~Gorbenko,$^{9}$ J.-O.~Hansen,$^{20}$ 
F.~W.~Hersman,$^{22}$ D.~W.~Higinbotham,$^{24}$ R.~Holmes,$^{18}$ 
C.~R.~Howell,$^{4}$ E.~Hughes,$^{1}$ B.~Humensky,$^{15}$ S.~Incerti,$^{19}$ 
C.W.~de Jager,$^{20}$ J.~S.~Jensen,$^{1}$ X.~Jiang,$^{17}$ 
C.~E.~Jones,$^{1}$ M.~Jones,$^{3}$ R.~Kahl,$^{18}$ H.~Kamada,$^{16}$ 
A. Kievsky,$^{5}$ I.~Kominis,$^{15}$ W.~Korsch,$^{21}$
K.~Kramer,$^{3}$ G.~Kumbartzki,$^{17}$ M.~Kuss,$^{20}$ 
E.~Lakuriqi,$^{19}$ M.~Liang,$^{20}$ N.~Liyanage,$^{20}$ J.~LeRose,$^{20}$ 
S.~Malov,$^{17}$ D.J.~Margaziotis,$^{2}$ J.~W.~Martin,$^{12}$ K.~McCormick,
$^{12}$
R.~D.~McKeown,$^{1}$ K.~McIlhany,$^{12}$ Z.-E.~Meziani,$^{19}$
R.~Michaels,$^{20}$ G.~W.~Miller,$^{15}$ J.~Mitchell,$^{20}$ S.~Nanda,$^{20}$ 
E.~Pace,$^{7,23}$ T.~Pavlin,$^{1}$ G.~G.~Petratos,$^{10}$ 
R.~I.~Pomatsalyuk,$^{9}$
D.~Pripstein,$^{1}$ D.~Prout,$^{10}$ R.~D.~Ransome,$^{17}$ Y.~Roblin,$^{11}$ 
M.~Rvachev,$^{12}$ A.~Saha,$^{20}$ G.~Salm\`{e},$^{6}$ M.~Schnee,$^{19}$ 
T.~Shin,$^{12}$ 
K.~Slifer,$^{19}$ P.~A.~Souder,$^{18}$ S.~Strauch,$^{17}$ R.~Suleiman,$^{10}$ 
M.~Sutter,$^{12}$ B.~Tipton,$^{12}$ L.~Todor,$^{14}$ M.~Viviani,$^{5}$ 
B.~Vlahovic,$^{13,20}$ 
J.~Watson,$^{10}$ C.~F.~Williamson,$^{10}$ H.~Wita{\l}a,$^{8}$ 
B.~Wojtsekhowski,$^{20}$ F.~Xiong,$^{12}$ J.~Yeh,$^{18}$ 
P.~\.{Z}o{\l}nierczuk$^{21}$}
\address{$^{1}$California Institute of Technology, Pasadena, CA 91125, USA\\
$^{2}$California State University, Los Angeles, Los Angeles, CA 90032, USA\\
$^{3}$College of William and Mary, Williamsburg, VA~23187, USA\\
$^{4}$Duke University, Durham, NC~27708, USA\\
$^{5}$INFN, Sezione di Pisa, Pisa, Italy \\
$^{6}$INFN, Sezione di Roma, I-00185 Rome, Italy \\
$^{7}$INFN, Sezione Tor Vergata, I-00133 Rome, Italy \\
$^{8}$Institute of Physics, Jagellonian University, PL-30059 Cracow, Poland\\ 
$^{9}$Kharkov Institute of Physics and Technology, Kharkov 310108, Ukraine\\
$^{10}$Kent State University, Kent, OH~44242, USA\\
$^{11}$LPC, Universit\'{e} Blaise Pascal, F-63177 Aubi\`{e}re, France\\
$^{12}$Massachusetts Institute of Technology, Cambridge, MA~02139, USA\\
$^{13}$North Carolina Central University, Durham, NC~27707, USA\\
$^{14}$Old Dominion University, Norfolk, VA~23508, USA\\
$^{15}$Princeton University, Princeton, NJ~08544, USA\\
$^{16}$Ruhr-Universit\"{a}t Bochum, D-44780 Bochum, Germany\\
$^{17}$Rutgers University, Piscataway, NJ~08855, USA\\
$^{18}$Syracuse University, Syracuse, NY~13244, USA\\
$^{19}$Temple University, Philadelphia, PA~19122, USA\\
$^{20}$Thomas Jefferson National Accelerator Facility, Newport News, 
VA 23606, USA\\
$^{21}$University of Kentucky, Lexington, KY~40506, USA\\
$^{22}$University of New Hampshire, Durham, NH~03824, USA\\
$^{23}$Dipartimento di Fisica, Universit\`a di Roma "Tor Vergata", Rome, Italy\\
$^{24}$University of Virginia, Charlottesville, VA~22903, USA}

\date{9 August 2002}
\maketitle

\begin{abstract}
A high precision measurement of 
the transverse spin-dependent asymmetry $A_{T'}$ in
$^3\vec{\rm He}(\vec{e},e')$ quasielastic scattering 
was performed in Hall A at
Jefferson Lab at values of the squared four-momentum transfer,
$Q^2$, between 0.1 and 0.6 (GeV/c)$^2$.
$A_{T'}$ is sensitive to the neutron magnetic form factor, $G_M^n$.
Values of $G_M^n$ at $Q^2 = 0.1$ and 0.2 (GeV/c)$^2$, extracted using
Faddeev calculations, were reported previously. 
Here, we report the extraction of $G_M^n$ for the remaining
$Q^2$-values in the range from 0.3 to 0.6 (GeV/c)$^2$ using a 
Plane-Wave Impulse Approximation calculation. The results
are in good agreement with recent precision data from
experiments using a deuterium target.
\end{abstract}



The electromagnetic form factors of the nucleon have been a
longstanding subject of interest in nuclear and particle physics.
They describe the distribution of charge and magnetization within
nucleons and allow sensitive tests of nucleon models based on
Quantum Chromodynamics. 
Precise knowledge of the form factors advances our understanding of
nucleon structure.

The proton electromagnetic form factors have been determined with
good precision at low values of the squared four-momentum transfer, $Q^2$,
while the neutron
form factors are known with much poorer precision 
because of the lack of free neutron targets.
Over the past decade, with the advent of high-quality 
polarized beams and targets, the precise determination of both the
neutron electric form factor, $G_E^n$, and the magnetic form factor,
$G_M^n$, has become a focus of experimental activity.  Considerable
attention has been devoted to the precise measurement of $G_M^n$.
While knowledge of $G_M^n$ is interesting in itself, it is also
required for the determination of $G_E^n$, which is usually measured
via the ratio $G_E^n/G_M^n$.  Furthermore, precise data for the nucleon
electromagnetic form factors are essential for the analysis of parity
violation experiments~\cite{sample,happex} 
designed to probe the strangeness content of the nucleon.

Until recently, most data on $G_M^n$ had been deduced from elastic and
quasi-elastic electron-deuteron scattering.  For inclusive
measurements, this procedure requires the separation of the longitudinal 
and transverse cross sections and the subsequent subtraction of a large
proton contribution. Thus, it suffers from large theoretical uncertainties
due in part to the deuteron model employed and in part to 
corrections for final-state
interactions (FSI) and meson-exchange currents (MEC).  
These complications can largely be avoided
if one measures the cross-section ratio of 
$d(e,e'n)$ to $d(e,e'p)$ at quasi-elastic kinematics.  Several recent
experiments \cite{Ankl94,Brui95,Ankl98,kubon2002} have employed this
technique to extract $G_M^n$ with uncertainties of 
$<$2\%~\cite{Ankl98,kubon2002} at $Q^2$ below 1 (GeV/c)$^2$.
Despite the high precision reported, however,
there is considerable disagreement among some of the
experiments~\cite{Mark93,Ankl94,Brui95,Ankl98,kubon2002} 
with respect to the absolute value of $G_M^n$.
The most recent deuterium data \cite{kubon2002} 
further emphasize this discrepancy. 

Thus, additional data on $G_M^n$, preferably obtained using a complementary
method, are highly desirable.
%
Inclusive quasi-elastic $^3\vec{\rm He}(\vec{e},e')$ scattering
provides just such an alternative approach \cite{Gao94}. In comparison to
deuterium experiments, this technique employs a different target and
relies on polarization degrees of freedom.  It is thus subject to
completely different systematics.  
As demonstrated recently \cite{xu2000}, 
a precision comparable to that of deuterium ratio experiments
can be achieved with the $^3{\rm He}$ technique.
 
The sensitivity of spin-dependent $^3\vec{\rm He}(\vec{e},e')$
scattering to neutron structure originates from the
cancellation of the proton spins in the dominant spatially symmetric
$S$ wave of the $^3$He ground state. As a result of this cancellation,
the spin of the $^3$He
nucleus is predominantly carried by the unpaired neutron 
alone \cite{BW84,friar90}.
Hence, the spin-dependent contributions to
the $^3\vec{\rm He}(\vec{e},e')$ cross section are expected to
be sensitive to neutron properties.
Formally, the spin-dependent part of the inclusive 
cross section is contained in two nuclear response 
functions, a transverse response $R_{T'}$ and a
longitudinal-transverse response $R_{TL'}$, which occur 
in addition to the
 spin-independent longitudinal and transverse responses $R_{L}$ and
$R_{T}$ \cite{twd86}. $R_{T'}$ and $R_{TL'}$ 
can be isolated experimentally by forming the spin-dependent asymmetry $A$ 
defined as
    $A = (\sigma^{h+}-\sigma^{h-})/(\sigma^{h+}+\sigma^{h-})$,
where $\sigma^{h^{\pm}}$ denotes the cross section for the two 
different helicities of the polarized electrons.
In terms of the nuclear response functions, $A$ can be 
written~\cite{twd86} 
\begin{equation}
\label{asym} 
A = \frac{-(\cos{\theta^{*}}\nu_{T'}R_{T'} +
  2\sin{\theta^{*}}\cos{\phi^{*}}\nu_{TL'}R_{TL'})}{\nu_{L}R_{L} +
  \nu_{T}R_{T}}
\end{equation}  
where the $\nu_{k}$ are kinematic factors and $\theta^{*}$ and
$\phi^{*}$ are the polar and azimuthal angles of target spin with
respect to the 3-momentum transfer vector ${\bf q}$. The response functions
$R_{k}$ depend on $Q^{2}$ and the electron energy 
transfer $\omega$. 
By choosing $\theta^\ast = 0$, {\it i.e.} by orienting the target
spin parallel to the momentum transfer ${\bf q}$, one selects the
transverse asymmetry $A_{T'}$ (proportional to $R_{T'}$).
Various detailed calculations \cite{ciofi,salme,rws93,ishi98,golak} have
confirmed that $R_{T'}$, and thus $A_{T'}$, is strongly sensitive to
$(G_M^n)^2$.


The experiment was carried out
in Hall A at the Thomas Jefferson National Accelerator Facility (JLab),
using a longitudinally polarized continuous-wave electron beam 
incident on a high-pressure
polarized $^{3}$He gas target \cite{jlabtarget}.  
Six kinematic points were measured corresponding to $Q^2 = 0.1$ to
$0.6$ (GeV/c)$^2$ in steps of 0.1 (GeV/c)$^2$.  An incident electron beam
energy, $E_i$, of 0.778 GeV was employed for the two lowest 
$Q^2$ values of the 
experiment, while the remaining points were obtained at $E_i = 1.727$ GeV.
The spectrometer settings of the six quasielastic kinematics are 
listed in Table I. 
To maximize
the sensitivity to $A_{T'}$, the target spin was oriented at $62.5^\circ$
to the right of the incident electron momentum direction.
This corresponds to $\theta^\ast$ from $-8.5^\circ$ to $6^\circ$, 
resulting in a contribution to
the asymmetry due to $R_{TL'}$ of less than $2\%$ at all kinematic settings,
as determined from plane-wave impulse approximation (PWIA) calculations.  
Further experimental details can be found in references [9,19,20].

\begin{table}
\begin{tabular}{|c|c|c|c|} 
 $Q^{2}$ (GeV/c)$^{2}$ & $E$ (GeV) & $E'$ (GeV)& $\theta$ (degree)\\
 \hline 
 0.10& 0.778& 0.717 &24.44\\ 
 0.193&0.778& 0.667 & 35.50\\
 0.30& 1.727& 1.559 & 19.21\\
0.40& 1.727& 1.506 & 22.62\\
0.50& 1.727& 1.453 & 25.80\\
0.60& 1.727& 1.399 & 28.85\\ 
\end{tabular}
\caption[]{The spectrometer settings for the six quasielastic 
kinematics of the experiment, where $E$ is the incident electron 
beam energy, $E'$ and $\theta$ are the 
spectrometer central momentum and scattering angle settings, respectively.}
\end{table}

\begin{figure}[t]
\psfig{file=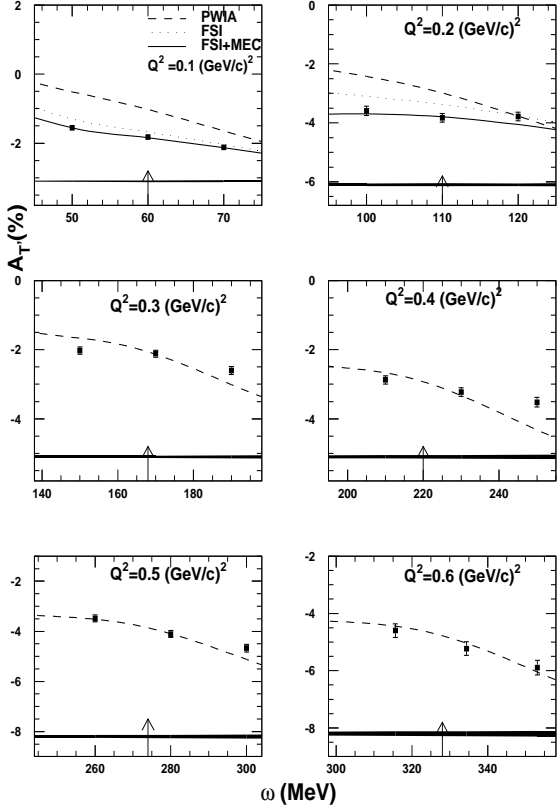,height=12.cm,width=8.cm}
\caption[]{The transverse asymmetry $A_{T'}$ near the peak of 
quasielastic scattering at the six kinematics of the experiment.
The data are shown with statistic uncertainties and
the experimental systematic uncertainties are shown as dark bands.  Also shown 
are the quasielastic peak positions.
The solid and the dash-dotted curves are 
the Faddeev calculation \protect\cite{golak} 
which includes the FSI and MEC effects and FSI effect only, respectively. 
The dashed curve is the PWIA calculation \protect\cite{salme}.} 
\label{atq1}
\end{figure}


Results for $A_{T'}$ (Fig.~\ref{atq1})
as a function of $\omega$ for all six kinematics of this experiment
together with the extracted $G^n_M$ values at the two lowest $Q^2$
kinematics of the experiment were reported previously \cite{xu2000}. 
A state-of-the-art non-relativistic Faddeev calculation 
\cite{golak2} was employed in the extraction 
of $G^n_M$ at these two $Q^2$ kinematics. 
As discussed in \cite{xu2000}, this calculation, while very 
accurate at low $Q^2$, is not believed to be sufficiently precise 
for a reliable extraction of $G_M^n$ from the $^3$He asymmetry data
at higher $Q^2$ because of its non-relativistic nature.
Thus, it was not used to extract $G^n_M$ for $Q^2$
values of 0.3 and 0.4 (GeV/c)$^2$, even though the Faddeev calculation
has been extended numerically to a $Q^2$ value up to 0.4 (GeV/c)$^2$. 
The high precision $^3$He quasielastic asymmetry data in the 
breakup region from the same experiment 
\cite{xiong} at $Q^2$ values of 0.1 and 0.2 (GeV/c)$^2$ 
provided stringent test of the Faddeev calculation and 
supported further the approach used in Ref. \cite{xu2000} in
extracting $G^n_M$ at $Q^2$ values of 0.1 and 0.2 (GeV/c)$^2$.
Thus, a fully relativistic three-body calculation is highly desirable
for a reliable extraction of $G^n_M$ at higher values of $Q^2$. Unfortunately,
such a calculation is not available and difficult to carry out at present time.

On the other hand, the size of 
FSI and MEC corrections to inclusive scattering data is well known to 
diminish with increasing momentum transfer, and so PWIA will likely
describe the data well at higher $Q^2$. Indeed as shown in Fig.~1, 
the PWIA \cite{salme} calculation provides excellent description 
of the data at $Q^2$ values of 0.5 and 0.6 (GeV/c)$^2$.
In light of this, we felt it was reasonable
to extract $G_M^n$ from our asymmetry data using PWIA. 
In order to estimate the model uncertainty of this procedure, 
we used results from the full Faddeev 
calculation up to a $Q^2$ value of 0.4 (GeV/c)$^2$ to study quantitatively
the size and $Q^2$-dependence of FSI and MEC corrections.

A recent PWIA calculation \cite{salme} which takes into account the 
relativistic kinematics and current 
using the AV18 NN interaction
potential and the H\"{o}hler nucleon form factor parameterization \cite{hoh}
(for the proton form factors and $G_E^n$) was used for the extraction 
of $G_M^n$ at $Q^2 \ge 0.3$ (GeV/c)$^2$.
In this calculation, the struck nucleon is described by a plane wave, and
the interaction between the nucleons in the spectator pair is 
treated exactly by including the NN and the Coulomb interaction 
between the pp pair. The de Forest CC1 off-shell 
prescription \cite{deforest} was adopted for 
the electron-nucleon cross section. Furthermore, the Urbana IX three-body 
forces \cite{threebody} were included in the $^3$He bound state.

To extract $G^n_M$, measured transverse asymmetry data from 
a 30 MeV region around the quasielastic peak were used.
The PWIA calculation \cite{salme} was employed to
generate $A_{T'}$ as a function of $G^n_M$ in the same 30 MeV-wide 
$\omega$ region. In doing so,
spectrometer acceptance effects were taken into account.
By comparing the measured asymmetries with the PWIA predictions, 
$G^n_M$ values could be extracted.
Results for $G_M^n$ were obtained in two ways: (a) by
taking the weighted average of $A_{T'}$ from three neighboring 
10 MeV bins around the quasi-elastic peak (30 MeV total for $\omega$)
and then extracting $G_M^n$ from this average asymmetry,
and (b) by first extracting $G_M^n$ from each these 10 MeV bins separately 
and then taking the weighted average of the resulting $G_M^n$ values.
Both methods yield essentially the same results (within 0.1\%).

The systematic uncertainty in $G_M^n$
is almost entirely due to
the systematic error from the determination 
of the beam and target polarizations, which is $1.7\%$ in $A_{T'}$ and
$0.85\%$ in $\delta G_M^n/G_M^n$.
Such a high precision in the determination of beam and 
target polarizations can be achieved by using elastic 
polarimetry \cite{xu2000}.
An additional systematic error occurs in the extraction of $G_M^n$
due the experimental uncertainty in the determination
of the energy transfer $\omega$.
The uncertainty due to this source is 1.4\% at 
$Q^2 = 0.3$ and becomes negligible ($< 0.5$\%) at the higher
$Q^2$ points.

The model uncertainty inherent in the extraction procedure
depends on the various ingredients of the calculation, such as
the NN potential, the proton nucleon form factors, relativity,
and the reaction mechanism, including FSI and MEC.
The main processes neglected in PWIA are FSI and MEC; therefore, these
two contributions are expected to dominate the overall model uncertainty.
As mentioned, we used results from the 
non-relativistic Faddeev calculation carried out up to a
$Q^2$ value of 0.4 (GeV/c)$^2$ to estimate the uncertainties resulting from
the omission of FSI and MEC. (Faddeev results for $Q^2 > 0.4$ 
were not generated because the calculation manifestly breaks
down in that kinematic regime)

To estimate the effect of FSI, the non-relativistic Faddeev calculation 
with FSI, corrected for relativistic effects, was compared 
\cite{thesis_xu,prc_long} with the relativistic PWIA calculation 
\cite{salme}.  Relativistic corrections to the Faddeev calculation were
derived from a comparison between the standard, relativistic PWIA 
calculation \cite{salme}
and a modified, non-relativistic PWIA calculation \cite{thesis_xu}.
One can thus study the size and the $Q^2$-dependence of the FSI effect
up to a $Q^{2}$ value of 0.4 (GeV/c)$^2$. 
As expected, FSI corrections to $A_{T'}$ decrease with increasing $Q^{2}$. 
The estimated errors in $A_{T'}$ due to the neglect of the FSI effect in PWIA 
are 9.0\%, 3.6\% for $Q^{2}$ of 0.3, 0.4, and on the order of 1-2 $\%$
for $Q^2$ values of 
0.5, and 0.6 (GeV/c)$^2$ based on an extrapolation 
beyond a $Q^2$ value of 0.4 (GeV/c)$^2$.

\begin{figure}
\psfig{file=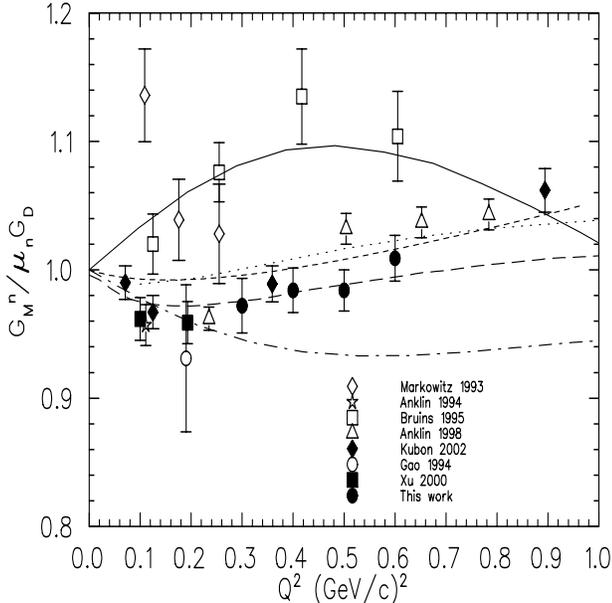,angle=90,height=8.cm,width=8.cm}
\caption[]{The neutron magnetic form factor $G_M^n$ in units of 
the standard dipole form factor $(1+Q^2/0.71)^{-2}$, at $Q^2$ values 
of 0.3 to 0.6 (GeV/c)$^2$ extracted using PWIA calculations. Also shown are
published measurements since 1990 and a few selected 
theoretical models. The Q$^2$
points of Anklin~94 \cite{Ankl94} and Gao~94 \cite{Gao94} 
have been shifted 
slightly for clarity. The solid curve is a recent cloudy bag model 
calculation \cite{lu}, the long dashed curve is a recent calculation 
based on a fit of the proton data using dispersion theory 
arguments \cite{mergell}, and 
the dotted curve is a recent analysis based on the vector meson 
dominance model \cite{lomon}. The dashed curve is a skyrme/soliton 
model calculation \cite{holzwarth}, and the dash-dotted curve is a 
relativistic quark model calculation \cite{schlumpf}.}
\label{fig:gmndatanew}
\end{figure} 

The MEC effect can be addressed in a similar manner.
Based on the Faddeev calculation \cite{golak}, we find that
MEC corrections to $A_{T'}$ near the top of the quasielastic peak
decrease exponentially as $Q^{2}$ increases. 
Similar conclusions have been drawn from studies of the quasi-elastic 
$\vec{d} (\vec{e}, e') $ process \cite{Arenhovel93}. 
We estimate the uncertainty due to the neglect of the MEC effect in PWIA
for $A_{T'}$ on top of the quasi-elastic peak to be
3.6\%, 2.4\%, 1.0\%, and 1.0\% for $Q^{2}$ of 
0.3, 0.4, 0.5, and 0.6 (GeV/c)$^2$, respectively.

The effect of various different off-shell prescriptions \cite{Caballero93}
was studied in the framework of the PWIA calculation,
and the contribution to the uncertainty of extracting
$G^n_M$ from $A_{T'}$ was found to be negligible.
Difference in $G_M^n$ arising from different choices of NN
                potential and other nucleon form factor
                parametrizations was
found to be about $1\%$.


Results for $G^{n}_M$ extracted at $Q^2 = 0.3$ to 0.6 (GeV/c)$^{2}$ using the 
PWIA calculation are presented in Table~\ref{tab:gmnresults} along
with statistical, systematic, and model uncertainties. 
The model uncertainties are obtained based on studies described 
previously, which may represent the lower limits only. 
The results are plotted in Fig~\ref{fig:gmndatanew} along with
the previously reported $G_M^n$ results \cite{xu2000} at 
$Q^2 = 0.1$ and 0.2 (GeV/c)$^2$, which were extracted 
using the Faddeev calculation. 
All other published results since 1990 are also shown.
The error bars shown on our data are the quadrature sum of the statistic 
and systematic uncertainties 
reported in Table~\ref{tab:gmnresults}, which do not 
include the estimated model uncertainty. 

While limitations exist in our approach, we note that our results
are in very good agreement with the recent deuterium ratio 
measurements from Mainz \cite{Ankl98,kubon2002}, and in disagreement 
with results by Bruins {\it et al.} \cite{Brui95}. 

In conclusion, we have measured the spin-dependent asymmetry $A_{T'}$
in the quasi-elastic $^3\vec{\rm He}(\vec{e},e')$ process with
high precision at $Q^2$-values from 0.1 to 0.6 (GeV/c)$^2$. 
In this Rapid Communication, we report the extraction of $G^n_M$ 
at $Q^2$ values of 0.3 to 0.6 (GeV/c)$^2$ based on PWIA calculations, which
are expected to be reasonably reliable in our range of $Q^2$.
We estimate the total uncertainty of our results to be about 4-6\%.
A more precise extraction of $G^n_M$ at these $Q^2$ values requires a 
fully relativistic three-body calculation, which is unavailable at present.
Efforts are underway to extend the theory into this regime 
\cite{glockle}.

\begin{table}
\begin{tabular}{|c|c|c|} 
 $Q^{2}$ (GeV/c)$^{2}$ & $G_M^n/G_M^n(Dipole)$ &Uncertainties
 (${\frac{\delta G^n_M}{G^n_M}}$)\\ \hline 
 0.30& 0.972&$\pm$0.014$\pm$0.016$^{+0.026}_{-0.054}$\\ 
 0.40& 0.984&$\pm$0.011$\pm$0.028$^{+0.028}_{-0.025}$\\ 
 0.50& 0.984&$\pm$0.009$\pm$0.024$^{+0.028}_{-0.013}$\\ 
 0.60& 1.010&$\pm$0.013$\pm$0.027$^{+0.031}_{-0.014}$\\ 
\end{tabular}
\caption[]{$G_M^n$ as a function of $Q^{2}$, the uncertainties are 
statistical, systematic and theoretical uncertainties, 
respectively.}
\label{tab:gmnresults}
\end{table}


We thank the Hall A technical staff and the Jefferson Lab
Accelerator Division for their outstanding support during this experiment.
This work was supported by the U.~S.~Department of Energy, DOE/EPSCoR,
the U.~S.~National Science Foundation, 
the Science and Technology Cooperation
Germany-Poland and the Polish Committee for Scientific Research, 
the Ministero dell'Universit\`{a} e della Ricerca
Scientifica e Tecnologica (Murst),
 the French Commissariat \`{a} l'\'{E}nergie Atomique,
Centre National de la
Recherche Scientifique (CNRS) and the Italian Istituto Nazionale di Fisica
Nucleare (INFN).
This work was supported by DOE contract DE-AC05-84ER40150
under which the Southeastern Universities Research Association
(SURA) operates the Thomas Jefferson National Accelerator Facility.
The numerical calculations were performed at the U.~S. 
National Energy Research Scientific Computer Center (NERSC) and
NIC in J\"{u}lich. 



\begin{references}
\bibitem{sample}B.~Mueller {\it et al.}, Phys. Rev. Lett. {\bf 78}, 3824 (1997). 
\bibitem{happex}K.A.~Aniol {\it et al.}, Phys. Rev. Lett. {\bf 82}, 1096 (1999).
\bibitem{Mark93} P.~Markowitz {\it et al.}, Phys. Rev. C {\bf 48}, R5 (1993).
\bibitem{Ankl94}H.~Anklin {\it et al.}, Phys. Lett. {\bf B336}, 313 (1994).
\bibitem{Brui95}E.E.W.~Bruins {\it et al.}, Phys. Rev. Lett. {\bf 75}, 21 
(1995).
\bibitem{Ankl98} H.~Anklin {\it et al.}, Phys. Lett. {\bf B428}, 248 (1998).
\bibitem{kubon2002}
G. Kubon, H. Anklin {\it et al.}, Phys. Lett. {\bf B524}, 26 (2002).
\bibitem{Gao94}H.~Gao {\it et al.}, Phys. Rev. C {\bf 50}, R546 (1994);
H.~Gao, Nucl. Phys. {\bf A631}, 170c (1998).
\bibitem{xu2000}
W. Xu {\it et al.}, Phys. Rev. Lett. 85, 2900 (2000).
\bibitem{BW84}B.~Blankleider and R.M.~Woloshyn, 
Phys. Rev. C {\bf 29}, 538 (1984).
\bibitem{friar90}J.L.~Friar {\it et al.}, Phys. Rev. C {\bf 42}, 2310 (1990).
\bibitem{twd86}T.W.~Donnelly and A.S.~Raskin, Ann. Phys. {\bf 169}, 247 (1986).
\bibitem{ciofi}
C. Ciofi degli Atti, E. Pace and
G. Salme' Phy. Rev. C {\bf 46}, R1591 (1992); 
Phy. Rev. C {\bf 51}, 1108 (1995). 
\bibitem{salme}A. Kievsky, E. Pace, G. Salm\`{e}, M. Viviani, 
Phys. Rev. C {\bf 56}, 64 (1997).
\bibitem{rws93}R.-W.~Schulze and P.~U.~Sauer, 
Phys. Rev. C {\bf 48}, 38 (1993). 
\bibitem{ishi98}S.~Ishikawa {\it et al.}, Phys. Rev. C {\bf 57}, 39 (1998).
\bibitem{golak}J.~Golak {\it et al.}, Phys. Rev. C {\bf 63}, 034006 (2001).

\bibitem{jlabtarget}J.S.~Jensen, Ph.D. Thesis, California Institute of 
Technology, 2000 (unpublished), available from \\
${\mbox{http://www.jlab.org/e94010/}}$;\\
P.L.~Anthony~{\it et al.}, Phys.~Rev.~D, {\bf 54}~6620~(1996).

\bibitem{thesis_xu}
W. Xu, Ph.D. thesis, Massachusetts Institute of Technology, 
2002 (unpublished). 
\bibitem{prc_long}
W. Xu, F.~Xiong {\it et al.}, to be submitted to Phys. Rev. C.


\bibitem{golak2} J.~Golak {\it et al.}, Phys. Rev. C {\bf 51}, 1638 (1995);
V.V. Kotlyer, H. Kamada, W. Gl\"{o}ckle, J. Golak,
Few-Body Syst. {\bf 28}, 35 (2000).


\bibitem{xiong}
F. Xiong {\it et al.}, Phys. Rev. Lett. {\bf 87}, 242501 (2001).
\bibitem{hoh}G.~H\"{o}hler {\it et al.}, Nucl. Phys. {\bf B114}, 505 (1976).

\bibitem{deforest}
T. de Forest {\it et al.}, Nucl. Phys. {\bf A392}, 232 (1983).

\bibitem{threebody}
S.C. Pieper, V.R. Pandharipande, R.B. Wiringa, J. Carlson, 
Phys. Rev. C {\bf 64}, 014001 (2001).


\bibitem{Arenhovel93} H. Arenh\mbox{\"{o}}vel {\it et al.}, 
Few-Body Syst. {\bf 15}, 109 (1993) 
\bibitem{Caballero93} J.A.~Caballero, T.W.~Donnelly and 
G.I.~Poulis. Nucl. Phys. {\bf A555} (1993)
\bibitem{lu}
D.H.~Lu, A.W.~Thomas, A.G.~Williams, Phys. Rev. C {\bf 57}, 2628 (1998). 
\bibitem{mergell}
P.~Mergell, U.-G. Mei{\ss}ner, D. Drechsel, Nucl. Phys. {\bf A596}, 367 (1996).
\bibitem{lomon}
E.L. Lomon, Phys. Rev. C {\bf 64}, 035204 (2001); nucl-th/0203081

\bibitem{holzwarth}
G. Holzwarth, Z. Phys. {\bf A356}, 339 (1996); hep-ph/0201138.

\bibitem{schlumpf}
F. Schlumpf, Phys. Rev. D {\bf 44}, 229 (1993).

\bibitem{glockle}
J. Golak, W. Gl\"{o}ckle, private communication.
\end{references}
\end{document}